\documentclass[aps,floatfix,preprint,preprintnumbers,nofootinbib,superscriptaddress,natbib]{revtex4}

\usepackage{graphicx,float}
\usepackage{epsfig}		
\usepackage{dcolumn}
\usepackage{bm}
\usepackage{subfigure}

\usepackage[all]{xy}
\usepackage{amsmath,upgreek}
\usepackage{amssymb}

\usepackage{pdfpages}
\usepackage{color}
\usepackage{graphicx,epstopdf}
\usepackage[colorlinks=true,
 linkcolor=red,
 urlcolor=purple,
 citecolor=blue,hyperindex]{hyperref}

\begin{document}

\title{Chaotic Behaviour of the Earth System in the Anthropocene}

\author{A. E. Bernardini}
\email{alexeb@ufscar.br}
\altaffiliation[On leave of absence from]{~Departamento de F\'{\i}sica, Universidade Federal de S\~ao Carlos, PO Box 676, 13565-905, S\~ao Carlos, SP, Brasil.}

\author{O. Bertolami}
\email{orfeu.bertolami@fc.up.pt}
\altaffiliation[Also at~]{Centro de F\'isica das Universidades do Minho e do Porto, Rua do Campo Alegre s/n, 4169-007, Porto, Portugal.}

\author{F. Francisco}
\email{frederico.francisco@fc.up.pt}
\affiliation{Departamento de F\'isica e Astronomia, Faculdade de Ci\^{e}ncias da
Universidade do Porto, Rua do Campo Alegre 687, 4169-007, Porto, Portugal.}

\begin{abstract}
It is shown that the Earth System (ES) can, due to the impact of human activities, exhibit chaotic behaviour. Our arguments are based on the assumption that the ES can be described by a Landau-Ginzburg model, which, in itself, predicts that the ES evolves through regular trajectories in phase space towards a Hothouse Earth scenario under a finite amount of human-driven impact. Furthermore, we find that the equilibrium point for temperature fluctuations can exhibit bifurcations and a chaotic pattern if human impact follows a logistic map. Our final analysis includes interactions between different terms of the planetary boundaries in order to gauge the predictability of our model.
\end{abstract}


\date{\today}
\maketitle

\renewcommand{\baselinestretch}{1.2}

\section{Introduction}

Recently, we have argued that the transition of the Earth System (ES) from the Holocene to other stable states is analogous to a phase transition, admitting a description by the Landau-Ginzburg (LG) Theory \cite{Bertolami:2018,Bertolami:2019,Barbosa:2020}. According to this approach, the relevant thermodynamic variable required to specify the state of the ES is the free energy, $F$. The phase transition is expressed in terms of an order parameter, $\psi$, which corresponds to a reduced temperature relative to the Holocene average temperature, $\langle T_{\rm H}\rangle$, i.e., $\psi = (T - \langle T_{\rm H}\rangle)/ \langle T_{\rm H}\rangle$.

In physics, the LG theory \cite{GB1,GB2} allows for describing the nonlinear evolution of small disturbances near a finite bifurcation of the free energy, driven by temperature fluctuations, $\psi$, from a stable to an unstable state of a thermodynamic system.
At the onset of such a bifurcation, the system becomes unstable at a critical value of $\psi$, i.e., $\psi_{\rm c} \neq 0$. Therefore, the LG theory \cite{GB3,GB4} models systems either in a highly stable state (in a state that lies deep in a low potential energy configuration valley), such that considerable energy is required to move them out of this stable state, or in an unstable state (at the top of a hill with a high potential energy configuration, where a small energy fluctuation can push them off the hill down towards a valley of lower potential energy). This is the qualitative picture (cf. Fig.2 presented in Ref.\cite{Steffen:2018}) used to describe possible (un)stable configurations of the ES in the Anthropocene.
Patterns from the LG theory can therefore mathematically map the phase-space-like trajectories of the ES, from the Holocene through the Anthropocene \cite{Steffen:2018}, with temperature fluctuations serving as a characteristic feature of macroscopic phase transitions\footnote{Analogous to the microscopic ones in the LG theory.}. Ref.~\cite{Steffen:2018} suggests that self-reinforcing feedbacks could push the ES towards a planetary threshold that, if crossed, would prevent the stabilisation of the climate at intermediate temperatures. This would lead to continuous warming of the ES towards a Hothouse Earth pathway, which could not be stopped even if anthropogenic greenhouse gas emissions were reduced. Quantitative tools and physical analogies for the parameterisation of such a phase transition threshold are therefore relevant for interpreting these ES dynamics. These tools are provided by the LG modelling discussed here.

Considering the great acceleration of the human activities visible from the second half of the 20th century onwards \cite{Steffen:2014}, the LG framework allows for obtaining the so-called Anthropocene equation (AE), i.e., the evolution equation of the ES describing the transition from the Holocene to the Anthropocene conditions. Such a physical approach enables the determination of the state and equilibrium conditions of the ES in terms of the driving physical variables, $\eta$ and $H$, where $\eta$ is associated with the astronomical, geophysical and internal dynamical drivers, and $H$ corresponds to human activities, introduced in the phase-transition model as an external field \cite{Bertolami:2019}.

In this work, we carry out a phase-space analysis of the temperature field, $\psi$, as previously examined in Ref.~\cite{Bertolami:2019}, under the condition that human activities follow a logistic map. Since this assumption is consistent with the hypothesis that human-driven changes are limited, the logistic map assumption does not affect our previous conclusions \cite{Bertolami:2019}, namely that the recently discussed Hothouse Earth scenario \cite{Steffen:2018} corresponds to an attractor of regular trajectories in the ES dynamical system. Naturally, the logistic map can lead to more complex behaviour, as depending on the intensity of human impact, it might give rise to stability point bifurcations and chaotic behaviour that precludes the prediction of the evolution of the ES temperature fluctuations.

The paper is organised as follows. In section \ref{newant}, we review the LG model proposal and discuss the AE, as well as the dynamical system emerging from this description in phase space. In section \ref{newchaos}, we introduce the hypothesis of the logistic map for human activities and demonstrate that it can lead to stability point bifurcations and chaotic behaviour in phase space. We also consider the interactions between different terms of the Planetary Boundaries (PB) \cite{Steffen:2015,Rockstrom:2009,Steffen:2011} in order to assess the predictability of our model. Finally, in section \ref{newconclusions}, we present our conclusions.

\section{The Anthropocene Equation phase-space \label{newant}}

Dynamical systems of physical interest can be typically described in terms of Lagrangian or Hamiltonian function, irrespectively of being classical, statistical or quantum. Regardless of the framework, the phase space variables can be evaluated.
The phase space of our model is fully specified through the variables $(\psi,\dot{\psi})$, once a set of initial conditions, $(\psi_0,\dot{\psi_0})$, is assumed. The initial value problem can be established and solved through the evolution equation, which leads to a trajectory of the dynamical system in the phase space.

In Refs.~\cite{Bertolami:2018,Bertolami:2019} it has been proposed that transitions of the ES are like phase transitions, which can be described by the LG Theory through the free energy function in terms of the order parameter $\psi$,
\begin{equation}
	F(\eta,H) = F_0 + a(\eta)\psi^2 + b(\eta)\psi^4 - \gamma H\psi, 
	\label{neweq:free_energy}
\end{equation}
where $F_0$ is a constant, the above mentioned set of natural parameters that affect the ES is denoted by $\eta$, with the natural effects modelled by $a(\eta)$ and $b(\eta)$, while the strength of the human activities, $H$, is set by $\gamma$.

Considering a general set of canonical coordinates, $q = (q_1,\,\dots,\,q_n)$, for the ES in our description, this should include, not only the order parameter $\psi$, but also the natural and human drivers, $\eta$ and $H$, respectively.
Therefore, we have $q=(\psi,\,\eta,\,H)$ \cite{Bertolami:2019}.
In this case, the Lagrangian function includes, besides the potential, which is identified with the free energy, a set of kinetic terms for the canonical coordinates,
{\setlength\arraycolsep{2pt}
\begin{eqnarray}
	\mathcal{L}(q,\dot{q}) = \frac{\mu}{2} \dot{\psi}^2 + \frac{\nu}{2} \dot{\eta}^2 - F_0 - a(\eta)\psi^2 - b(\eta)\psi^4 + \gamma H\psi,
\end{eqnarray}}
where $\mu$ and $\nu$ are constants.

As we focus our analysis on the Anthropocene, the ES is dominated by the effects of human activities. Hence, the effect of these are greater and faster than the longer time scales of natural features \cite{Bertolami:2019}. Therefore, in order to study the ES currently, the term in $\dot{\eta}^2$ can be safely dropped\footnote{Of course, we could have introduced a kinetic term for the human activities, $\dot{H}^2$. As explained in Ref.~\cite{Bertolami:2019}, this term is associated with quite fast effects of the human activities on themselves, and even though this feedback loop is present, we assume instead that $H$ is an external force and drop its kinetic term.}.

This reduces the system to a single canonical coordinate, the order parameter $\psi$. Following the Hamiltonian approach, we can identify the canonical conjugate momentum:
\begin{equation}
	p = \frac{\partial \mathcal{L}}{\partial \dot{\psi}} = \mu \dot{\psi},
\end{equation}
and the following Hamiltonian,
\begin{equation}
	\mathcal{H}(\psi, p) = \frac{p^2}{2\mu} + a \psi^2 + b \psi^4 - \gamma H\psi,
\end{equation}
from which follows the Hamilton equations,
\begin{equation}
	\dot{\psi} = \frac{\partial \mathcal{H}}{\partial p}, \quad \dot{p} = - \frac{\partial \mathcal{H}}{\partial \psi},
\end{equation}
that provide the evolution equations of the dynamical system.

\begin{figure}
	\centering
		\includegraphics[width=\columnwidth]{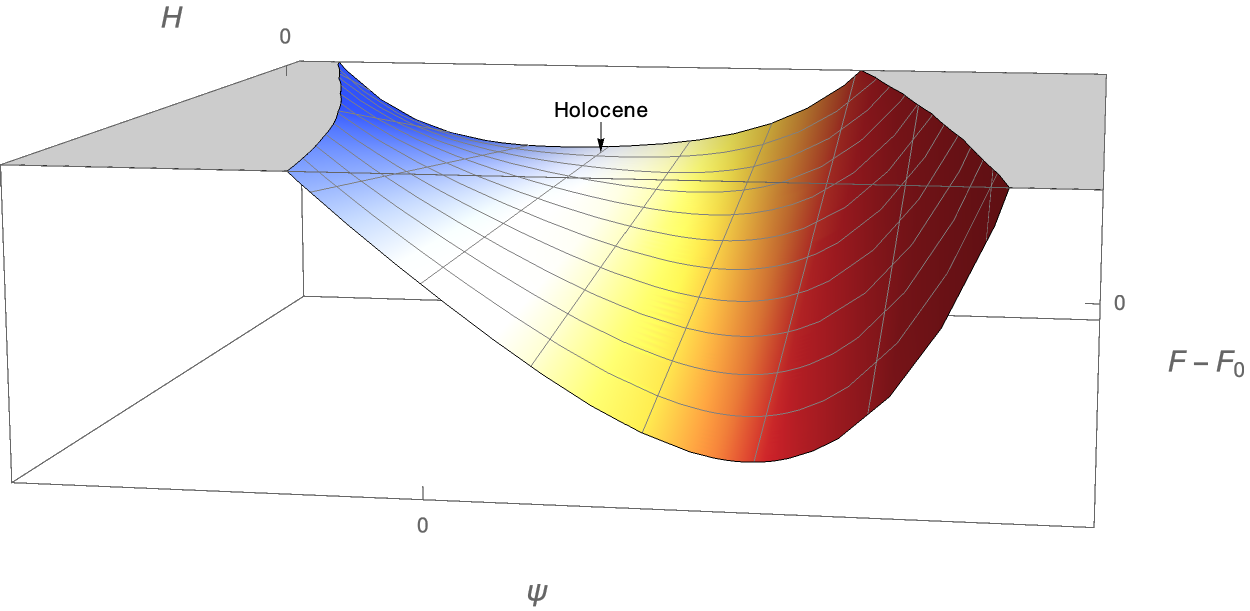}
	\caption{Stability landscape of the ES in terms of $\psi$ and $H$. Figure from Ref.~\cite{Bertolami:2019}.}
	\label{newfig:stability_landsacpe}
\end{figure}

With these equations, the phase portrait of the dynamical system can be obtained, and the corresponding orbits and attractors can be identified. The stability landscape of the temperature field, $\psi$, is depicted in Fig.~\ref{newfig:stability_landsacpe}, from which the Holocene minimum described in Ref.\,\cite{Bertolami:2018} can be identified at the center of the valley, for $H=0$. By shifting values of $H$ away from $H=0$, one clearly sees that the human intervention opens up a deeper hotter minimum, the Hothouse Earth \cite{Steffen:2018}, which in fact corresponds to a dynamical attractor \cite{Bertolami:2019}.

At the Anthropocene, $\eta$ is fixed and $H$ is treated as a parameter with a time dependence. The ES dynamics is reflected in the phase space and, thus, on the position and strength of its attractors, as exemplified in Ref.~\cite{Bertolami:2019}.
In that case, for $b \simeq 0$, the cubic term  can be dropped in the equations of motion, which now correspond to a harmonic oscillator with an external force, $H(t)$. 
Then, before considering more realistic scenarios for the external force, the simple case of a linear time evolution can be described by \cite{Bertolami:2019}:
\begin{equation}
	H(t) = H_0 t.
	\label{neweq:linear} 
\end{equation}
With this, the equations of motion are given by \cite{Bertolami:2019}:
\begin{equation}
	\mu \ddot{\psi} = - 2 a \psi + \gamma H_0 t,
\end{equation}
and the departure from equilibrium, $\dot{\psi}(0)=0$, with ${\psi}(0)=\psi_0$, by the analytical solution,
\begin{equation}
	\psi(t) = \psi_0 \cos (\omega t) + \alpha t,
\end{equation}
where $\omega = \sqrt{2a / \mu} $ is an angular frequency, $\alpha = \gamma H_0 / 2 a$.

As discussed in Ref.~\cite{Bertolami:2019}, this solution corresponds to elliptical trajectories in the phase space with moving foci of the form
\begin{equation}
	{{\Psi}^2 \over \psi_0^2 } + {\dot{\Psi}^2 \over \psi_0^2 \omega^2} = 1,
\end{equation}
where $\Psi = \psi + \alpha t$. 

These results provide a qualitative picture that holds even after introducing the cubic term $\propto \psi^3$ into the free energy expression, Eq.~\eqref{neweq:free_energy}. Its effect is to slightly deform the elliptic orbits and to slow its movement towards higher values of $\psi$. In fact, the equations of motion can be solved numerically, for instance, and the trajectory of the ES can be obtained for any set of initial conditions \cite{Bertolami:2019}. This analysis can be generalised for any time dependence of $H(t)$ and, of course, the system is only stable if $H(t)$ is bounded -- an issue that will be considered in the following discussion.

\section{Logistic map and chaotic behaviour in the Anthropocene}
\label{newchaos}

The linear model for the evolution of human activities, Eq. (\ref{neweq:linear}), presented in the previous section, is intended only as a simple example to illustrate the process of obtaining the trajectory of the ES in its phase space. The next natural step is to attempt a more realistic assumption regarding $H$.

A reasonable incremental step would be to consider an exponential evolution, following the pattern observed in many socio-economic trends over the past century, in what has come to be known as The Great Acceleration \cite{Steffen:2014}.

However, the growth of all human (and natural) activities is ultimately constrained by the available resources within the ecosystem. Given the global spread of humanity, the ecosystem to consider is therefore global, but still limited. Even when accounting for gains in efficiency brought about by technology, the fundamental fact remains that resources and the capacity to replenish them are ultimately finite.

Considering resource limitation as a reasonable hypothesis, a suitable function to describe the evolution of human activities is the logistic map. Indeed, the logistic map is widely used in ecological problems on the dynamics of populations. Precisely due to environmental limitations, the rate of growth of a population is constrained and will ultimately reach a saturation value, corresponding to the equilibrium between a given species, its predators, and the resources available in its environment \cite{LV1,LV2,Arditi89,Berryman92,Kingsland95}.

The logistic map is a relatively simple yet highly rich model, exhibiting features such as stable attractors, bifurcations, and chaotic behaviour, depending on the values of a single parameter \cite{MayMap76,Jakobson81,Eckmann81}.

In general, the existence and stability of equilibrium points for dynamical systems can be stablished through the Lyapunov theorem. A critical point $(\psi_{\rm c}, p_{\rm c})$ is said to be \emph{Lyapunov stable} if any trajectory starting at a given neighbourhood of that point remains within a finite neighbourhood of $(\psi_{\rm c}, p_{\rm c})$. In the continuous domain of the time variable, $(\psi_{\rm c},\,p_{\rm c})$ is identified by solving the constraint $(\dot{\psi},\, \dot{p}) = (0,\,0)$, which results into $p_{\rm c} = 0$ and $\psi_{\rm c}$ as the unique real solution of the cubic equation,
\begin{equation}\label{newNovo}
	\mathcal{O}(\psi) = 2 a \psi + 4 b \psi^3 = \gamma H.
\end{equation}

Still within the context of the LG framework introduced in the previous section, it is important to emphasise that periods of significantly hotter conditions, such as the Paleocene-Eocene Thermal Maximum (PETM), were characterised by transient states rather than stabilised Hothouse Earth conditions. Feedback mechanisms, such as chemical weathering and biogeochemical buffering, drove the ES back towards cooler states. In light of Eq.~\eqref{newNovo}, this would correspond to critical values, $(\psi_{\rm c}, p_{\rm c})$, for which $H = 0$, and the natural drivers $\eta$, implicitly affecting $a \equiv a(\eta)$ and $b \equiv b(\eta)$, would dominate.
Therefore, either stable or unstable equilibrium Hothouse Earth conditions could arise, for instance, from overwhelming natural regulatory mechanisms during the evolution of the ES. This can be considered the ``natural'' condition of the ES, which does not require adaptation for natural-only forcings, as discussed in the previous section. Anthropogenic forcing, driven by the coupling between $H$ and $\psi$, mediated by $\gamma$, changes the ES from the ``natural dynamics'' to the Anthropocene, as suggested in the previous section for linear growth in $H$. This shifts the equilibrium and stability patterns, which, through Eq.~\eqref{newNovo}, now depend on the balance between anthropogenic and natural forcings.

Hence, it is now necessary to specify the human drivers, collectively denoted by $H$. As suggested in Ref.~\cite{Bertolami:2019}, a fruitful strategy is to consider the impact of the human activities in the context of the PB Framework \cite{Steffen:2015}, in which the state of the ES is specified through a set of 9 parameters, such that the human impact can be measured in terms of the actual magnitude of these parameters as compared to their values at the Holocene. The later set of values is usually referred to as Safe Operating Space (SOS) \cite{Rockstrom:2009}.

The most general form of $H$ given by the impact of the human activities on the PB parameters, $h_i$, contains also the interactions among them \cite{Bertolami:2019}:
\begin{equation}
	H = \sum_{i=1}^{N} h_i + \sum_{i,j=1}^{N} g_{ij} h_i h_j + \ldots,
	\label{neweqn:human_action}
\end{equation}
with $N=9$, if we are considering the PB Framework. A mathematically convenient analysis requires that $[g_{ij}]$ be read as a $9 \times 9$ symmetric and non-degenerate matrix, with $g_{ij} = g_{ji}$ and $\det[g_{ij}] \neq 0$, respectively.
For such a 9-variable PB Framework, the 9th variable could be associated with the ``Technosphere''. In particular, if $h_1 > 0$, $h_{9} < 0$ and $g_{1\,9} > 0$, the destabilising effect of $h_1$ can be mitigated by Technosphere contributions \cite{Bertolami:2019}, which would maintain the equilibrium point due to human activities closer to its Holocene value \cite{Bertolami:2018}.
The second order coupling terms involving $h_i h_j$ represent the interactions among the various effects of the human intervention on the ES. 
As discussed in Ref.\,\cite{Bertolami:2019}, they can affect the equilibrium configuration and suggest some mitigation strategies depending on the sign of the matrix entries, $g_{ij}$, and their strength \cite{Bertolami:2019}. This methodology was considered to show that the interaction term between the climate change variable ($\rm CO_2$ concentration), $h_1$, and the oceans acidity, $h_2$, is non-vanishing and on the order of $10\%$ of the value of the single contributions by themselves \cite{Barbosa:2020}.
Finally, suppressed higher order interaction terms are sub-dominant and their importance has to be established empirically. 
Therefore, our analysis shall be restricted to second order contributions and, in fact, to a subset of PB.  

Of course, the above considerations also assume that the $h_i$ terms do not depend on the temperature. However, it is most likely, in physical terms, that $h_i = h_i(\psi)$, meaning in fact that the human action does affect the free energy of the ES \cite{Bertolami:2019}.
The actual evolution of the temperature can be estimated as the associated equilibrium state which evolves as $\langle \psi \rangle \sim H^{1 \over 3}$ \cite{Bertolami:2018}. In this case, considering the beginning of the Anthropocene \cite{Jones:2004} about 70 years ago, and assuming that the growth of $H$ remains linear \cite{Bertolami:2019}, if the effect of the human action does lead to an increase of $1\,{\rm K}$ since then, one should expect in 2050 a temperature increase of $\sqrt[3]{2} = 1.26\,{\rm K}$.
Hence, a quite general conclusion is that the critical point of the dynamical system corresponds to an ES trajectory evolving towards a minimum of the free energy where the temperature is greater than $\langle T_H\rangle$.

In fact, a global temperature, $T$, somewhat larger than the Holocene averaged temperature, $\langle T_H \rangle$, could lead to a chain failure of the main regulatory ecosystems of the ES, for which tipping point features have already been detected \cite{Steffen:2018}. Therefore, predicting the behaviour of the ES accurately in order to engender mechanisms that force their associated phase space trajectories to remain close to the Holocene minimum is essential.

In such a context, we notice that the stability pattern can drastically change if the function $H$ describing the human intervention receives contributions from the PB parameters that are constrained by some kind of discretised logistic growth rate. Let us than assume that at least one of the PB parameters, $h_i$, iteratively evolves according to the recurrence relation of a logistic map,
\begin{equation}
	h_{1(n+1)} = r\, h_{1(n)}(1 - \alpha^{-1}\,h_{1(n)}).
\end{equation}
where $i=1$ has been arbitrarily chosen.

The logistic map is strictly related to a logistic function which describes the growth rates and their corresponding level of saturation through the evaluation of the logistic equation,
\begin{equation}
	\dot{h}_{1} = \nu\, h_{1}(1 - \kappa^{-1} h_{1}),
\end{equation}
where the constant $\nu$ defines the growth rate and $\kappa$ its carrying capacity.

\begin{figure}[h!]
	\includegraphics[width=0.8\columnwidth]{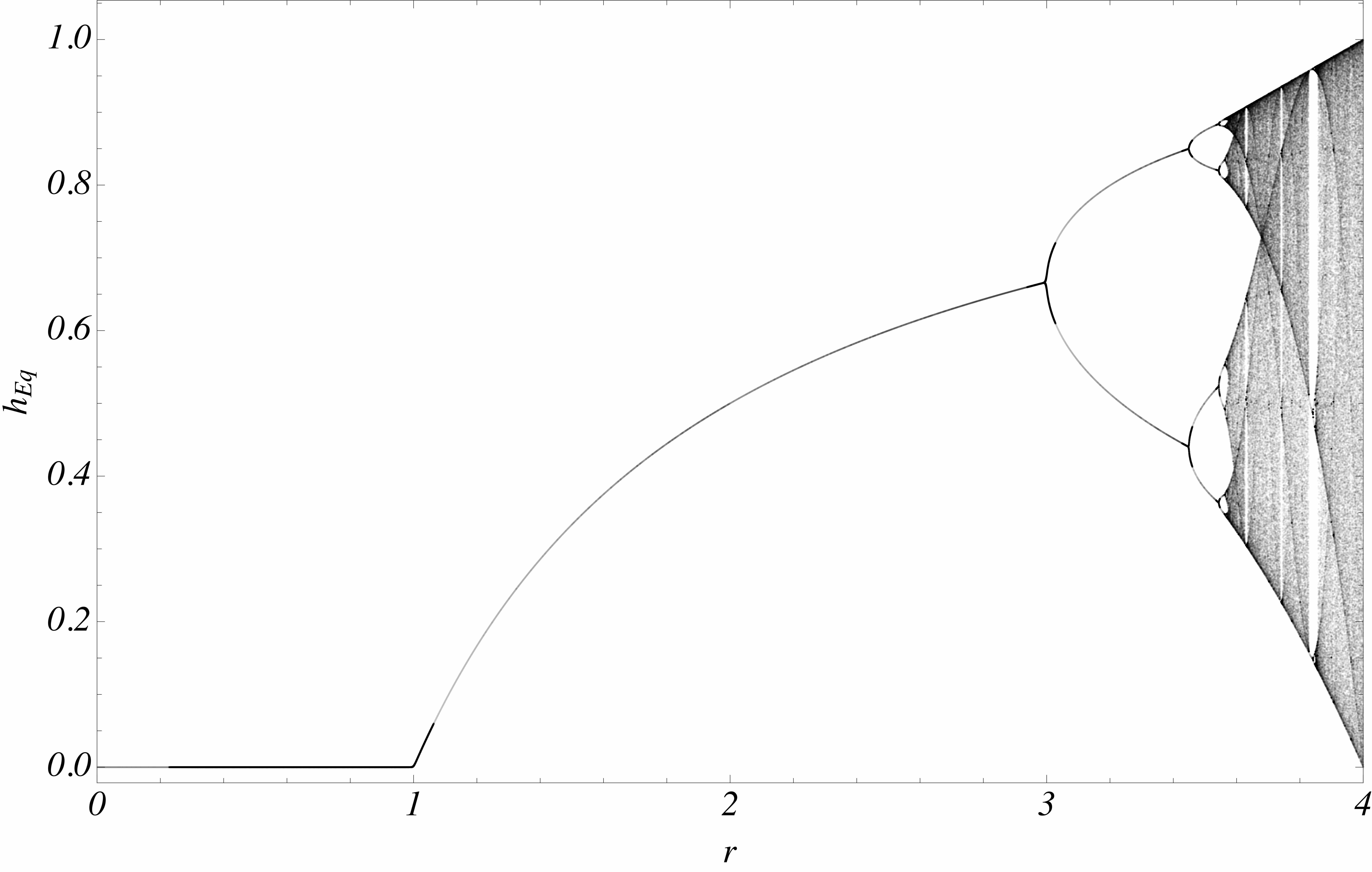}
	\includegraphics[width=0.8\columnwidth]{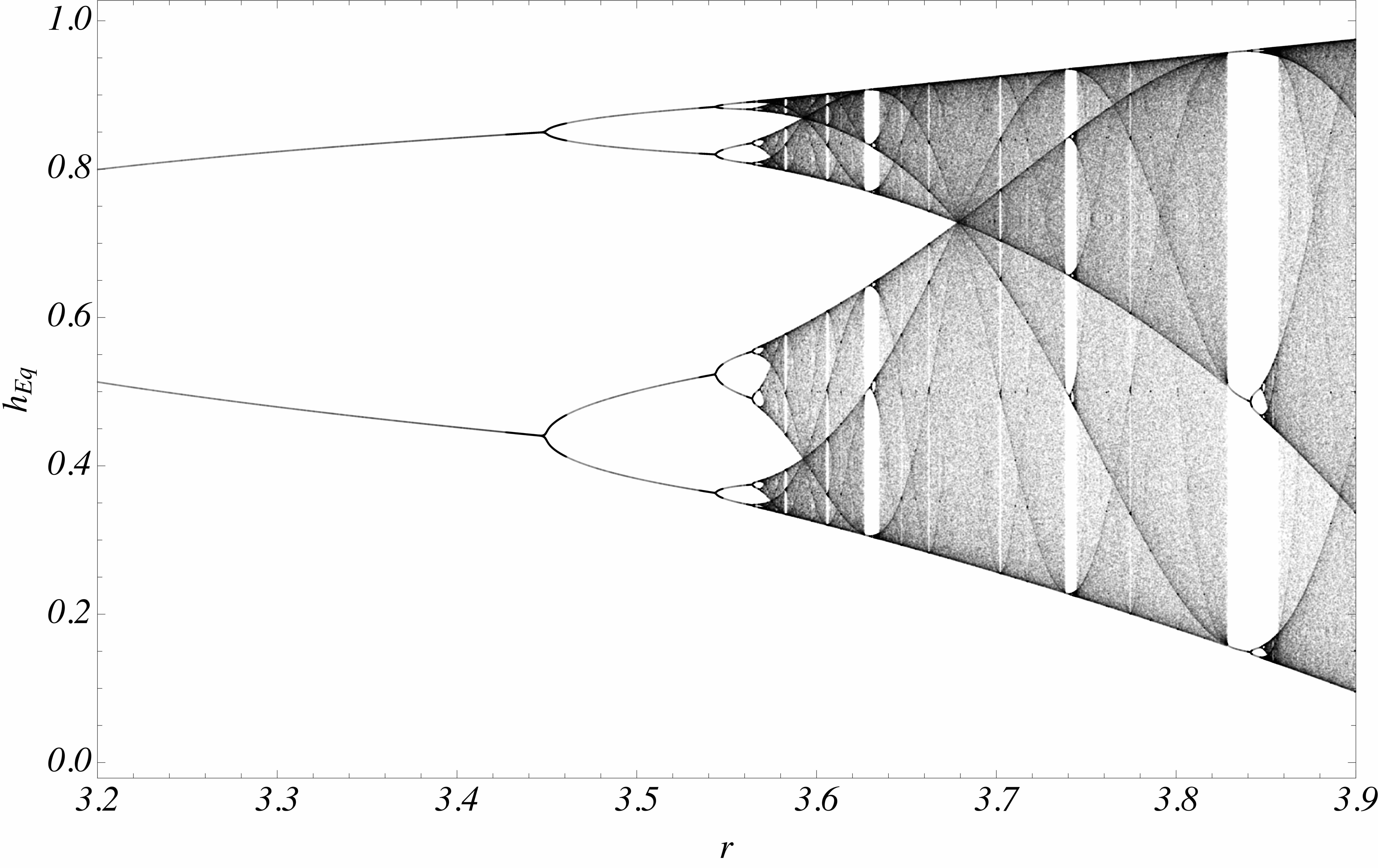}
	\caption{Bifurcation diagram for the logistic map ($\alpha = 1$).}
	\label{newChao01}
\end{figure}

Instabilities and chaotic patterns emerge from the conversion of a dynamical (time) variable continuous domain into a discrete domain, where the typical time scale between the steps of the discrete chain is a year, setting the map correspondence to: 
\begin{eqnarray}
	\dot{h}_{1} &\mapsto& h_{1(n+1)} - h_{1(n)},\nonumber\\
	\nu &\mapsto& r-1,\nonumber\\
	\kappa &\mapsto& \alpha(1-r^{-1}),\nonumber
\end{eqnarray}
The logistic map, even for $\alpha = 1$, as depicted in Fig.~\ref{newChao01}, provides a good illustration of how equilibrium point bifurcations and chaotic behaviour can arise from a simple set of non-linear dynamical equations.

The interpretation of Fig.~\ref{newChao01} is well-known as it shows all possible behaviours of the equilibrium points, ${h}_{1}\equiv h^{Eq}$, as a function of $r$: $i)$ for $r$ between $0$ and $1$, the impact on $h^{Eq}$ vanishes, independent of the initial conditions; $ii)$ for $r$ between $1$ and $2$, $h^{Eq}$ quickly approaches the value $r-1/ r$, independent of the initial conditions; $iii)$ for $r$ between $2$ and $3$, $h^{Eq}$ eventually approaches the same value of $r-1/r$, but it slightly fluctuates around that value for $r \lesssim 3$; $iv)$ from $r$ = 3 to $ r=1 + \sqrt{6}\approx3.4495$, a bifurcation pattern arises, drastically changing the dynamical evolution of $h^{Eq}$, which approaches permanent oscillations between two equilibrium values given by $h^{Eq}_{\pm}=\left(r+1\pm {\sqrt {(r-3)(r+1)}}\right)/2r$; $v)$ for $r$ between $\sim 3.4495$ and $\sim 3.5441$ the bifurcations are doubly degenerated, the lengths of the $r$ parameter intervals that duplicate the number of equilibrium points decreases rapidly producing a kind of period-doubling cascade with the ratio between the lengths of two successive bifurcation intervals approaching the Feigenbaum constant $\delta \sim 4.6692$ \cite{Feig78}. That is the chaotic dependence on the boundary value problem of the corresponding dynamical variable $h^{Eq}$. In what follows we shall see that this logistic map has a direct impact on the equilibrium temperature predictions, $\psi^{Eq}$. 

\begin{figure}[h!]
	\includegraphics[width=0.8\columnwidth]{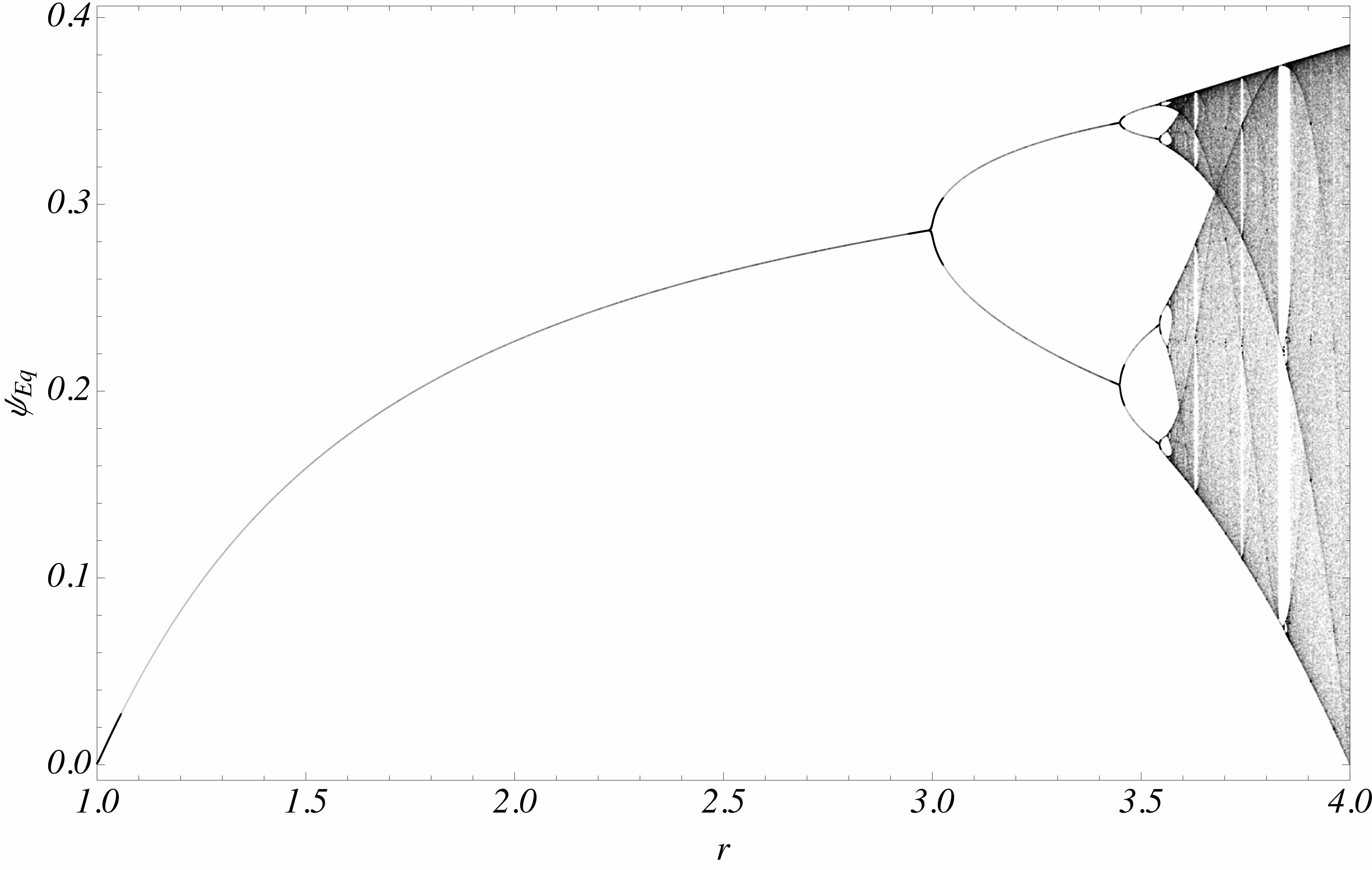}
	\caption{Bifurcation diagram for $\psi^{Eq}$ for $\mathcal{O}(\psi) = h_1$, with $a=b=1$.}
	\label{newChao02}
\end{figure}

From Eq.~\eqref{newNovo}, with $H\sim h_{1}$, a preliminary conclusion is that the equilibrium point, $\psi^{Eq}$, can be drastically affected by the $h_1$ growth rate $r$, since an equivalent map for $\psi^{Eq}\mapsto \psi^{Eq}_n$ can be identified by
\begin{equation}\label{newNovo2}
	\psi^{Eq}_n \mapsto \psi^{Eq}_{n+1} = \psi^{Eq}_{n+1}(\psi^{Eq}_n)
\end{equation}
such that, through the operator $\mathcal{O}$ identified from Eq.~\eqref{newNovo},
\begin{eqnarray}
	\hspace{-.8cm}\psi^{Eq}_{n+1}(\psi^{Eq}_n)=&&\hspace{-.6cm}\ \mathcal{O}^{-1}\left[H(h_{1(n+1)})\right]\nonumber\\
	=&&\hspace{-.6cm}\ \mathcal{O}^{-1}\left[
	H\left(r\, h_{1(n)}\left(1 - \alpha^{-1}\,h_{1(n)}\right)\right)
	\right] \nonumber\\
	=&&\hspace{-.6cm}\ \mathcal{O}^{-1}
	\left[H
	\left(
	r\, H^{-1}
	\left(\mathcal{O}[\psi^{Eq}_n]\right)\right.\right.\nonumber\\
	&&\left.\left.\qquad\left(1 - \alpha^{-1}H^{-1} \left(\mathcal{O}[\psi^{Eq}_n]\right) \right)
	\right)
	\right],\qquad\qquad
	\label{newNovo3B}
\end{eqnarray}
which exhibits a similar bifurcation diagram as depicted in Fig.~\ref{newChao02}.
It is interesting to notice the presence of some {\em islands of unstable equilibrium domains}, where the bifurcation pattern can be macroscopically identified.

Aiming to get from the above scenario a more realistic phenomenological description, we consider a couple of parameters, $h_1$ and $h_2$, and assume, in particular, that the second parameter is conditioned by the first one\footnote{The choice of $h_1$ and $h_2$ is essentially methodological. As one notices in the final discussion, the model can be complexified by the inclusion of additional degrees of freedom, from $h_1$ to $h_9$. Otherwise, the logistic growth dynamics are not applicable without anthropogenic forcing. In fact, some level of unpredictability in the chaotic dynamics (not driven by the logistic map) might still emerge from natural feedbacks (e.g., ice-albedo interactions, methane release) which, however, are out of the scope of the modelling here considered.}.

Parametrising the coupling between $h_1$ and $h_2$ by an interaction term given by $ g_{12}\,h_1 h_2$, such a reduced ES can be described by a normalised human activity function, $\tilde{H}$, written as
\begin{equation}
	\tilde{H}= \frac{H}{H_T} = \frac{h_1+h_2 + g_{12}\,h_1 h_2 }{h_{1T}+h_{2T} + g_{12}\,h_{1T} h_{2T}},
	\label{newHHHH}
\end{equation}
where constants $h_{1(2)T}$ were introduced to establish a consistent comparative analysis.

\begin{figure}[h!]
	\includegraphics[width=.8\columnwidth]{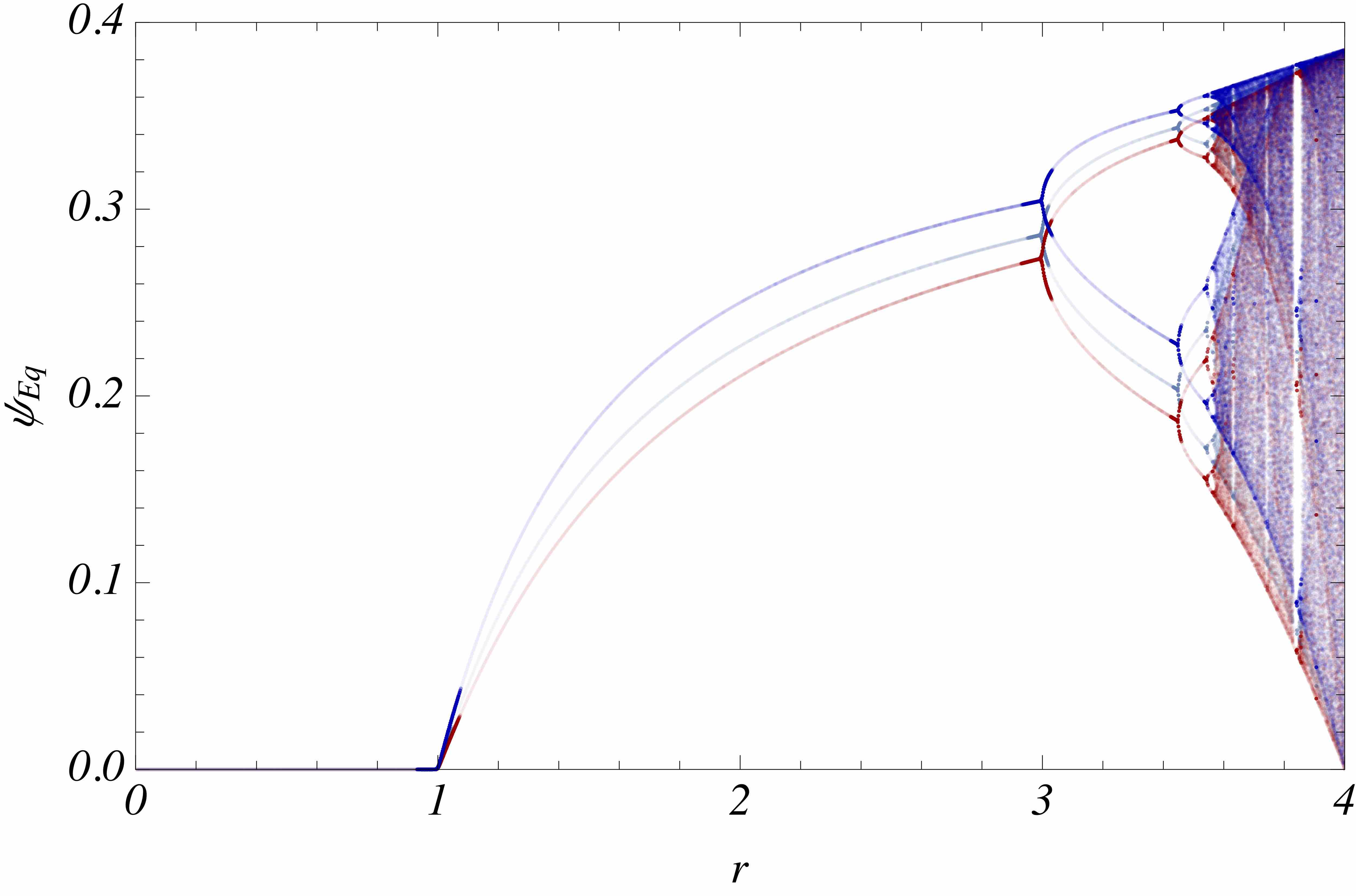}
	\includegraphics[width=.8\columnwidth]{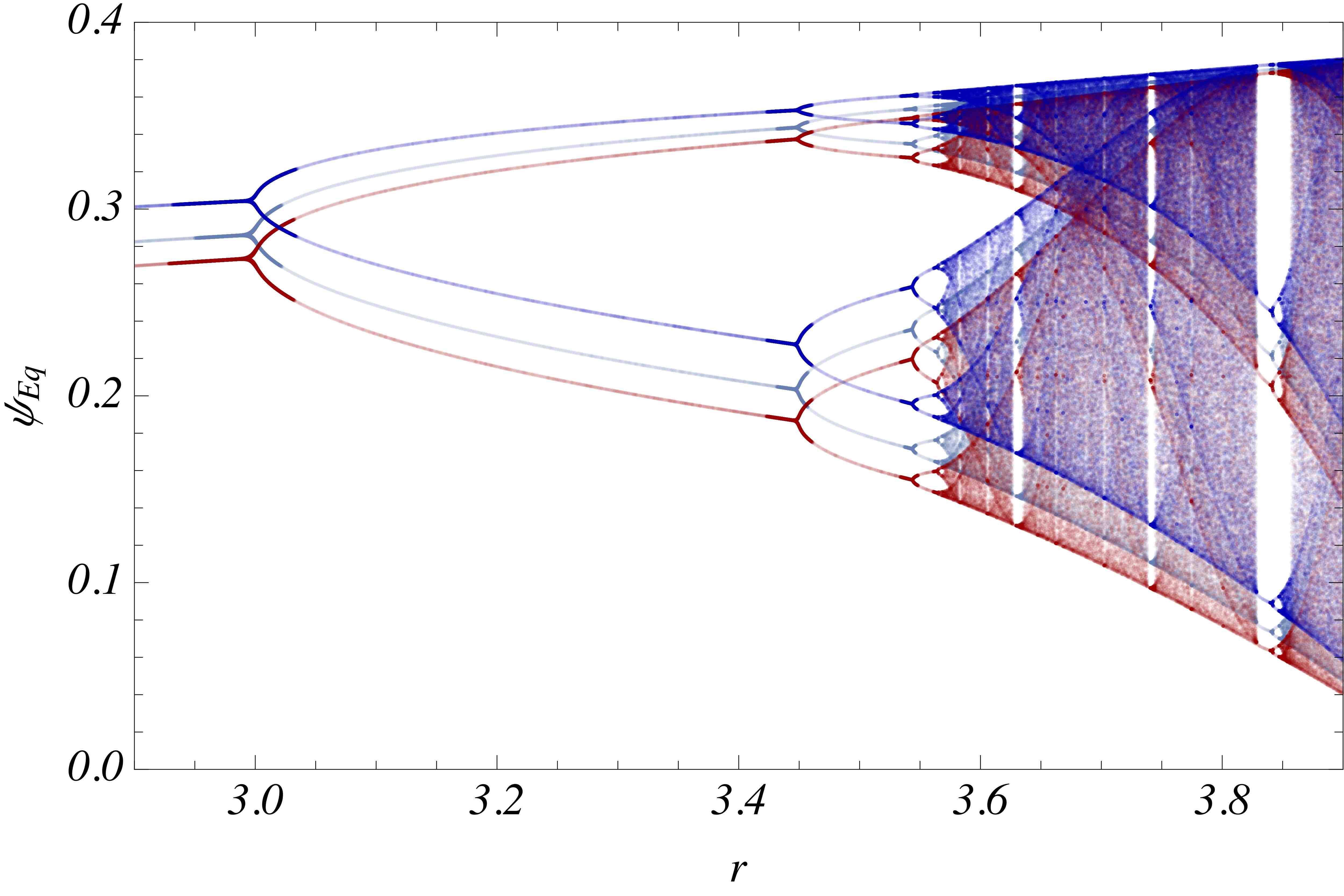}
	\caption{Bifurcation diagram for $\psi^{Eq}$ in case of a linear coupling between, parametrised by $\mathcal{O}(\psi) = \tilde{H} = (h_1 + \omega h_1^2)/({1+\omega})$. Results are for $\omega= 0.2$ (red), $0$ (black) and $-0.2$ (blue), with $a=b=1$. They are presented in a {\em zoom in view} in the second plot.}
	\label{newChao03}
\end{figure}

For the hypothesis of a linear correlation between $h_1$ and $h_2$, that is, $h_2 = \lambda h_1$, one has
\begin{equation}
	\tilde{H} = \frac{h_1 + \omega h_1^2}{1+\omega}, \qquad \mbox{with} \quad \omega = \frac{g_{12}\,\lambda}{1+\lambda},
\end{equation}
and from Eq.~\eqref{newNovo3}, the map for $\psi^{Eq}$, is given by
\begin{equation}\label{newNovo3}
	\psi^{Eq}_{n+1}(\psi^{Eq}_n) = \mathcal{O}^{-1}
	\left[
	\frac{r}{1+\omega}
	\left(
	x_n(1-x_n) + r \omega x_n^2(1-x_n)^2
	\right)
	\right]
\end{equation}
with
\begin{equation}\label{newNovo3C}
	x_n = \frac{1}{2\omega}\left(\left(1+4\omega(1+\omega)\mathcal{O}[\psi^{Eq}_{n}] \right)^{1/2}-1\right),
\end{equation}
for which the bifurcation diagram is depicted in Fig.~\ref{newChao03}. We point out that the coupling, $\omega$, just modulates the amplitude of the critical point, $\psi^{Eq}$, not affecting the above mentioned {\em islands of equilibrium}, which correspond to the emerging blanks within the chaotic pattern.

It is worth pointing out that the presence of bifurcations and chaotic pattern in the behaviour of ES drastically affects the capability to make predictions and to control the behaviour of the ES temperature. One could say that these features are to be expected given the complexity of the ES and its interactions, however, our model allows for depicting as well as quantifying the approach to this complex behaviour via the tracking of the actual value of the parameter $r$, which, in principle, can be extracted from data on the human activity. 

The above results admit two implementations depending on the coupling between $h_1$ and $h_2$:

\vspace{0.3 cm}
\noindent
{\em{Scenario 1 -- Coupling between $h_1$ and $h_2$ evolving with the same growth rate, $r$, and different saturation points, $\alpha = 1$ and $\alpha > 1$, respectively.}}

Considering two unconstrained dynamical variables, $h_1$ and $h_2$, this case corresponds to a couple of logistic maps identified by
\begin{eqnarray}\label{newNovo01A}
	h_1 &\mapsto& h_{1(n+1)} = r\, h_{1(n)}(1 - h_{1(n)});\\
	h_2 &\mapsto& h_{2(n+1)} = r\, h_{2(n)}(1 - \alpha^{-1}\,h_{2(n)}).
\end{eqnarray}

\begin{figure}
	\hspace{-.25cm}
	\includegraphics[width=.5 \columnwidth]{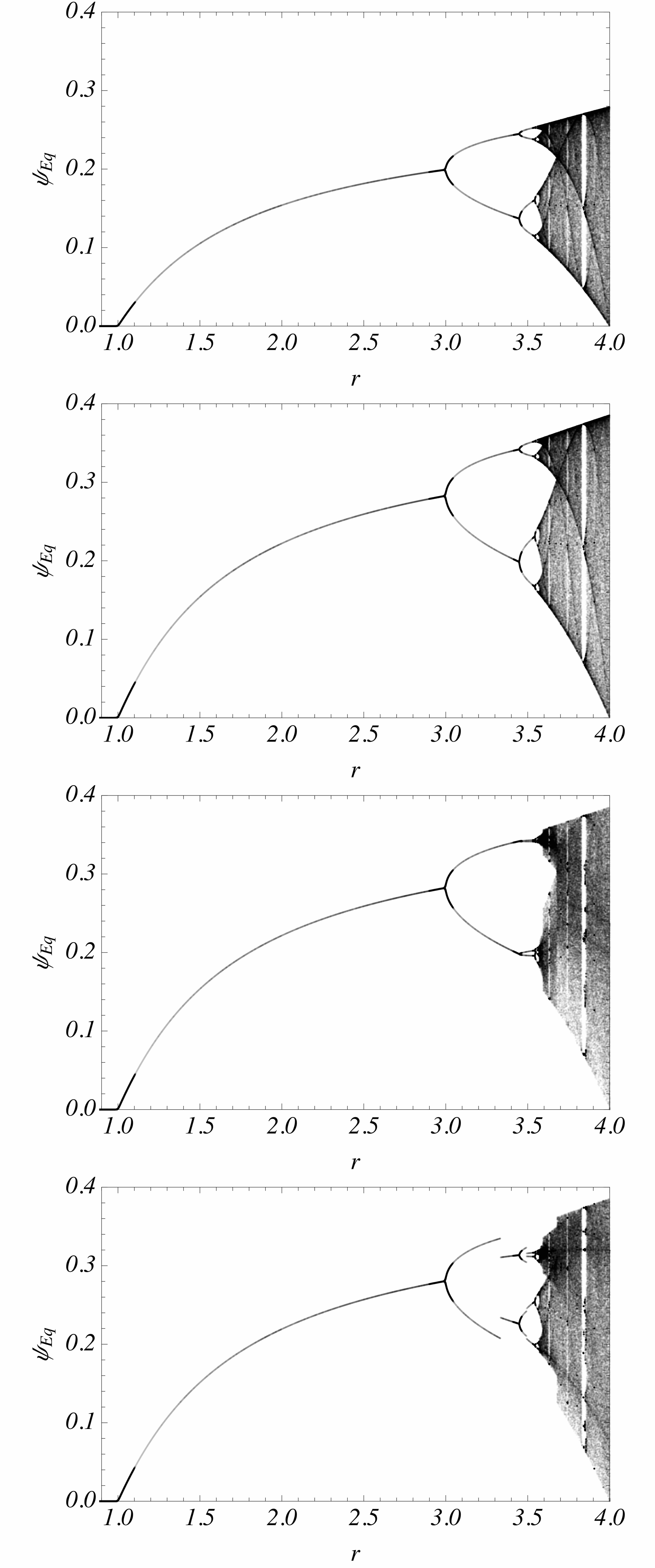}
	\hspace{-.25cm}
	\includegraphics[width=.5 \columnwidth]{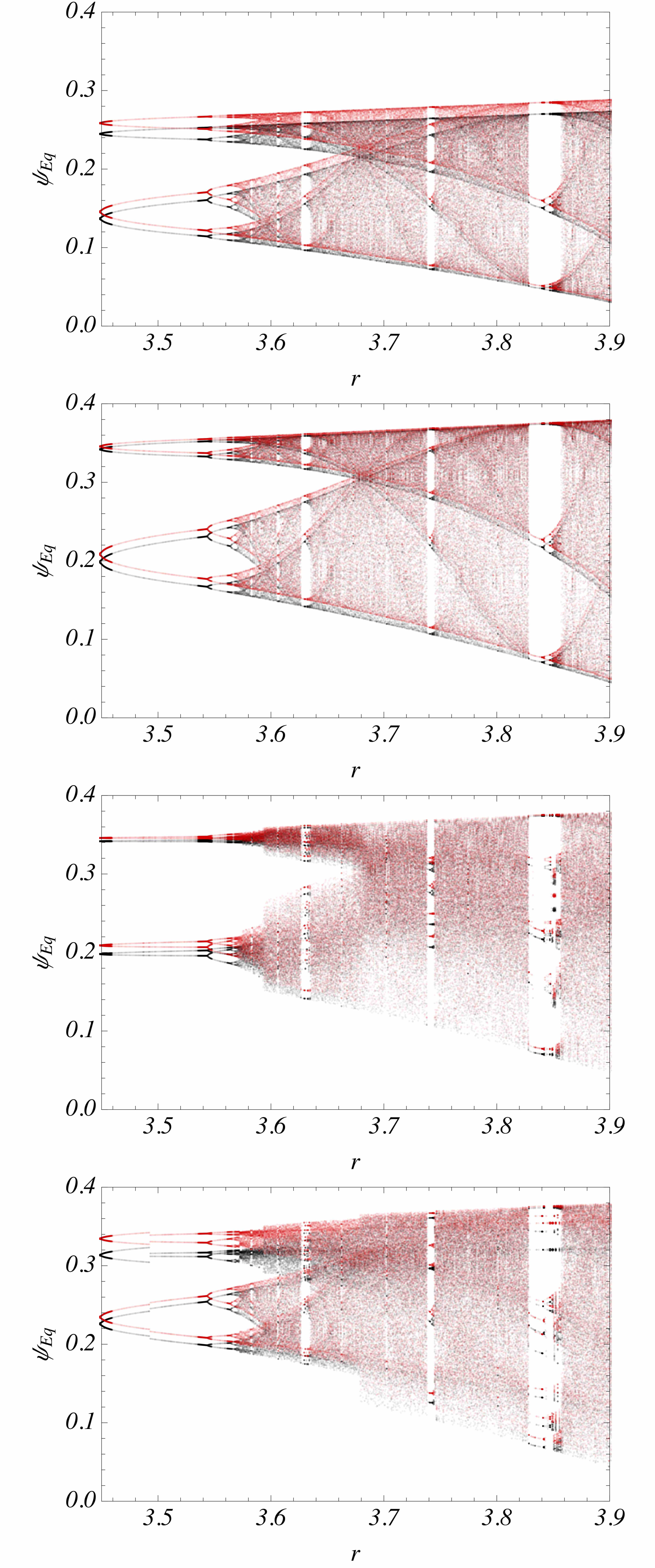}
	\caption{Maps for $\psi^{Eq}$ as function of $r$ for the Scenario 1, for $g_{12} = 0.1$. Results are for $\alpha = 1$ (first row), $1.25$ (second row), $5$ (third row), and $10$ (forth row). Second column depicts the results in {\em zoom view} for $g_{12} = 0.1$ (black) and $-0.1$ (red). The case $g_{12} = 0$ corresponds to the intermediate configuration.}
	\label{new001}
\end{figure}

Hence, for the human activity function expressed by Eq.~\eqref{newHHHH}, the map for $\psi^{Eq}$ as function of $r$ is depicted in Fig.~\ref{new001}, for $g_{12} = 0.1$ and $-0.1$.

In particular, the analysis of the continuous domain relating $h_1$ to $h_2$, shows that
\begin{equation}
	\frac{dh_1}{dh_2} = \frac{\nu\,h_1(1-h_1)}{\nu\,h_2(1-\kappa^{-1} h_2)} = \frac{h_1(1-h_1)}{h_2(1-\kappa^{-1} h_2)},
\end{equation}
which leads to the following constraint,
\begin{equation}
	h_2 = \frac{c\,h_1}{h_1(c \kappa^{-1} - 1) + 1}.
\end{equation}
For an integration constant, $c$, set equal to $\kappa$, one has $h_2 = {\kappa\,h_1}$, a linear relation in a continuous domain that can be mapped into a discrete domain as
\begin{eqnarray} \label{newNovo01}
	h_{1(n)} &\mapsto& h_{2(n)}=\alpha\, h_{1(n)};\nonumber\\
	h_{1(n+1)} &\mapsto& h_{2(n+1)}=\alpha\, h_{1(n+1)}.
\end{eqnarray}
From the linear map Eq.~\eqref{newNovo01} one notices that
\begin{eqnarray}\label{newNovo01A}
	h_1 &\mapsto& h_{1(n+1)} = r\, h_{1(n)}(1 - h_{1(n)}),
\end{eqnarray}
consistently returns
\begin{eqnarray}\label{newNovo01A}
	h_2 &\mapsto& h_{2(n+1)} = r\, h_{2(n)}(1 - \alpha^{-1}\,h_{2(n)}).
\end{eqnarray}
That is, the linear correspondence between $h_1$ and $h_2$ is just a particular solution of the {\em Scenario 1}.

\vspace{0.3 cm}
\noindent
{\em{Scenario 2 -- Coupling between $h_1$ and $h_2$ evolving with the same saturation points (set equal to unity) with different growth rates, $r$ and $\alpha r$ ($\alpha < 1$).}}

Considering two unconstrained dynamical variables, $h_1$ and $h_2$, it ensues two distinct logistic maps:
\begin{eqnarray}\label{newNovo01A}
	h_1 &\mapsto& h_{1(n+1)} = r\, h_{1(n)}(1 - h_{1(n)});\nonumber\\
	h_2 &\mapsto& h_{2(n+1)} = r\,\alpha\, h_{2(n)}(1 - h_{2(n)}).
\end{eqnarray}

\begin{figure}
	\hspace{-.25cm}
	\includegraphics[width=.5\columnwidth]{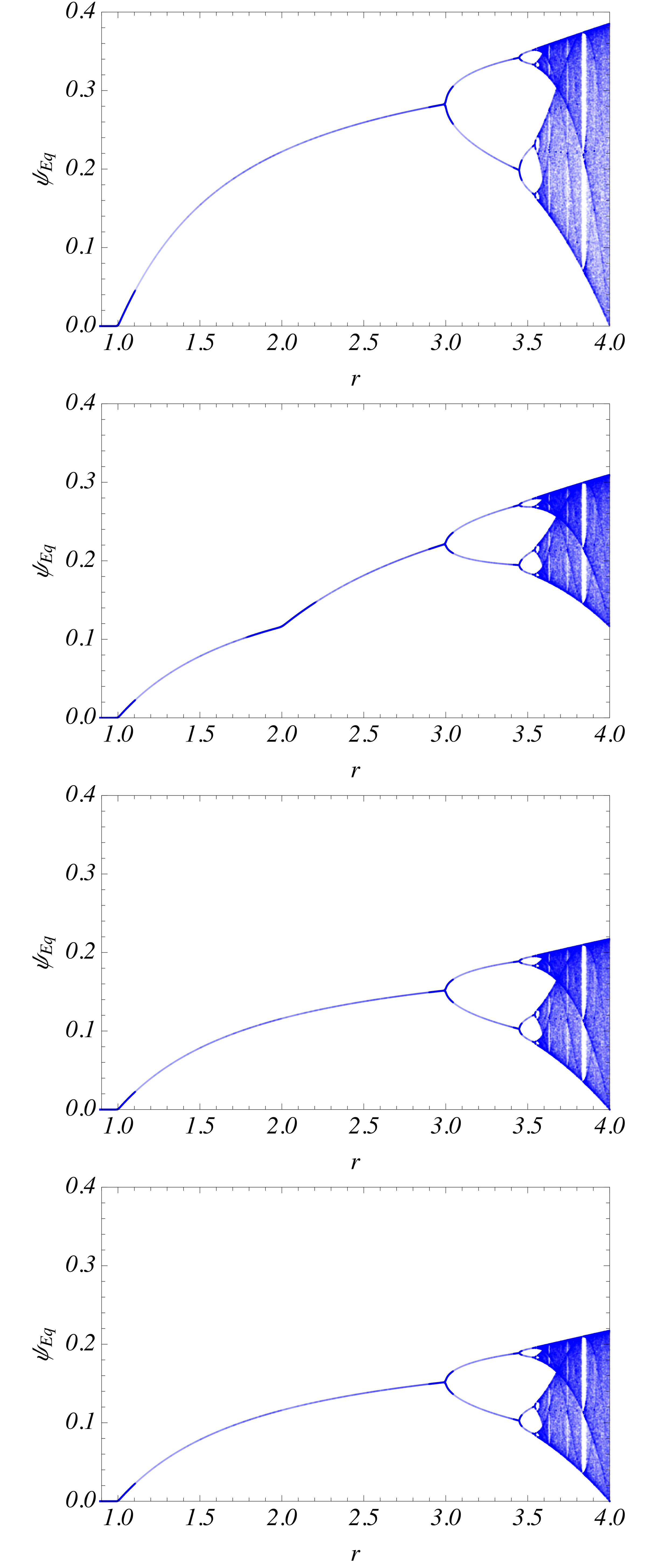}
	\hspace{-.25cm}
	\includegraphics[width=.5\columnwidth]{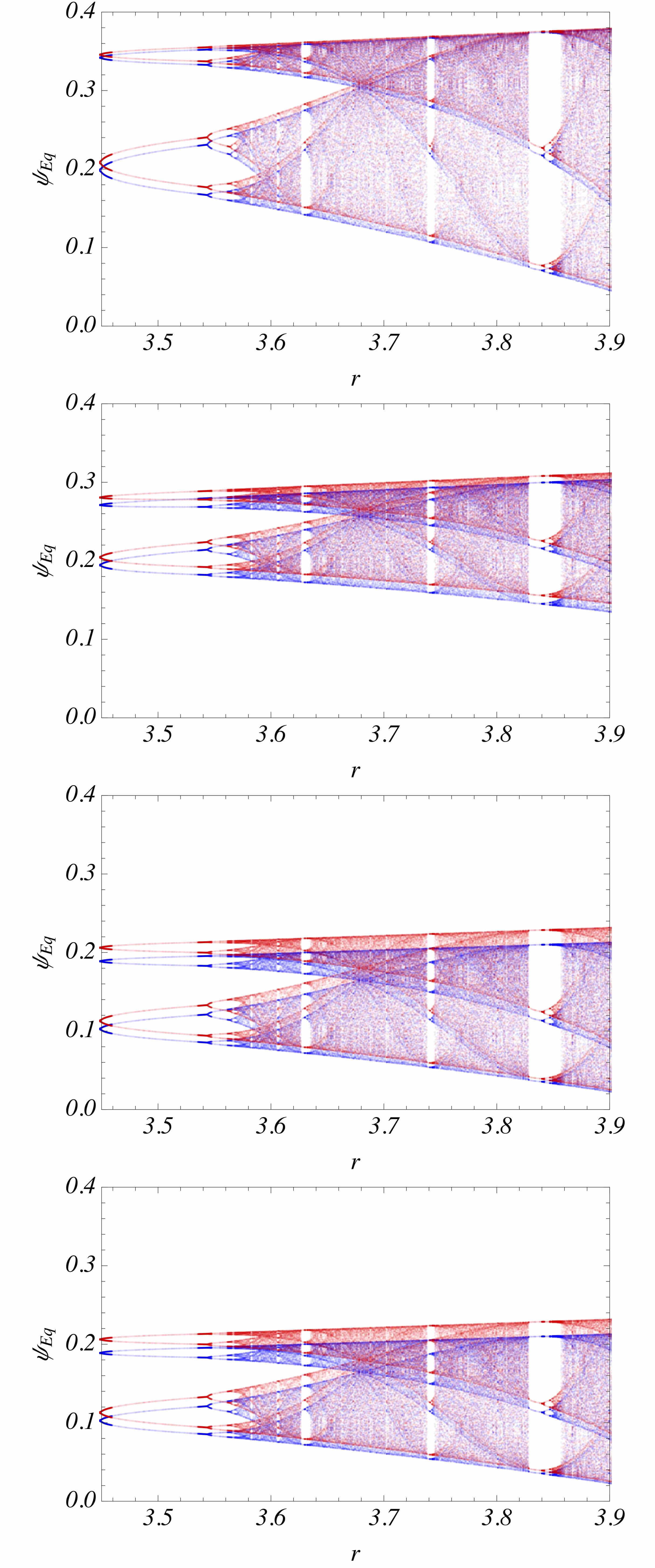}
	\caption{Maps for $\psi^{Eq}$ as function of $r$ for the Scenario 2, for $g_{12} = 0.1$. Results are for $\alpha = 2$ (first row), $1$ (second row), $0.8$ (third row), and $0.2$ (forth row). Second column depicts the results in {\em zoom view} for $g_{12} = 0.1$ (blue) and $-0.1$ (red). The case $g_{12} = 0$ corresponds to the intermediate configuration.}
	\label{new002}
\end{figure}

Hence, for the human activity function again expressed by Eq.~\eqref{newHHHH}, then the map for $\psi^{Eq}$ as function of $r$ is depicted in Fig.~\ref{new002}, for $g_{12} = 0.1$ and $-0.1$.

As in the previous case, an analysis of the continuous domain relating $h_1$ to $h_2$, shows that
\begin{equation}
	\frac{dh_1}{dh_2} = \frac{\nu\,h_1(1-h_1)}{\nu\,\kappa\,h_2(1- h_2)} = \frac{h_1(1-h_1)}{\kappa\,h_2(1-h_2)},
	\end{equation}
	which leads to the following constraint,
	\begin{equation}
	h_2 = \frac{h_1^{\kappa}}{c\,(1-h_1)^{\kappa} + h_1^{\kappa}},
\end{equation}
where $c$ is an arbitrary integration constant. The above result shows that {\em Scenarios 1 and 2} are topologically distinct.

From Figs.~\ref{new001} and \ref{new002} one notices that {\em Scenarios 1 and 2} exhibit bifurcation patterns with paired discontinuities at zeroth and first order derivatives, respectively. This drastically affects the chaotic pattern. However, the $r$ intervals from which equilibrium points, $\psi^{Eq}$, are accountable, can be evaluated and are still clearly identifiable in both cases. 

\vspace{0.3 cm}
\noindent
{\em{General phenomenological approach}}

We show here that the above results will ensue from a more general setup. Assuming a larger set of human activity contributions, $h_i$, from the logistic map parametrisation it follows that:
\begin{equation}
	h_{i(n+1)} = r \alpha_i h_{i(n)}(1-\beta_i h_{i(n)}),
\end{equation}
which are distinct from each other just by their growth rates and saturation points, $r \alpha_i$ and $\beta_i$, i.e. by denoting
\begin{equation}
	h_{i} \mapsto x_{n+1} = r \alpha_i x_n(1-\beta_i x_n),
\end{equation}
the $H$ behaviour can be reflected by a truncated version of Eq.~\eqref{neweqn:human_action}, 
\begin{eqnarray}
	\hspace{-1cm}
	H_{n+1} 	&=& \sum_{i=1}^{9} h_{i(n+1)} + \sum_{i,j=1}^{9} g_{ij} h_{i(n+1)} h_{j(n+1)}\nonumber\\
 			 	&=& \sum_{i=1}^{9} r \alpha_i x_n(1-\beta_i x_n) \nonumber\\
				&&+ \sum_{i,j=1}^{9} g_{ij} r^2 \alpha_i \alpha_j x_n x_n(1-\beta_i x_n)(1-\beta_j x_n)\nonumber\\
				&=& A\,x_n - B\,x_n^2 - C\,x_n^3 + D\,x_n^4,
	\label{neweqn:human_action2}
\end{eqnarray}
with
\begin{eqnarray}
	A 	&=& r\,\sum_{i=1}^{9} \alpha_i > 0,\nonumber\\
	B 	&=& r\,\sum_{i=1}^{9} \alpha_i \beta_i - r^2 \sum_{i,j=1}^{9} g_{ij}\alpha_i \alpha_j,\nonumber\\
	C 	&=& r^2 \sum_{i,j=1}^{9} g_{ij}\alpha_i \alpha_j (\beta_i + \beta_j),\nonumber\\
	D 	&=& r^2 \sum_{i,j=1}^{9} g_{ij}\alpha_i \alpha_j \beta_i \beta_j ,\nonumber
\label{neweqn:human_action3}
\end{eqnarray}
that is a polynomial map of forth order driven by four phenomenological parameters, $A,\,B,\,C$, and $D$, which can also be constrained by $Max\{H\} = H_T$, so that there remains just three free parameters. Despite the inherent complexity, it is expected that this model, likewise those discussed above, affects the corresponding (multidimensional) chaotic attractor in a similar way as identified for the equilibrium point $\psi^{Eq}$ in the two parameter analysis.

\section{Discussion and Outlook}
\label{newconclusions}

In this work we have considered a description of the ES in terms of the LG Theory of phase transitions and investigated whether some of the variables accounting for human activities can be modelled by a logistic map.
We have shown that it leads to a rather rich array of possible trajectories of the ES in the Anthropocene, including regular and predictable trajectories, as well as bifurcations and chaotic behaviour.
Furthermore, as previously discussed \cite{Bertolami:2019,Barbosa:2020}, the interaction among different PB parameters was investigated.

ES Models (ESM) are fundamental tools for predicting and simulating interactions among Earth's physical, chemical, biological, ecological, and human systems. These models integrate components of the climate system, such as the atmosphere, oceans, land surface, and ice, along with biogeochemical processes (e.g., carbon and nitrogen cycles) and, in some cases, human influences such as land use and emissions.
They are typically structured by a set of partial differential equations that describe fluid dynamics, radiation transfer, and thermodynamic processes across various components of the ES. The models represent a range of feedback loops and nonlinear interactions, such as the greenhouse effect, ocean-atmosphere exchanges, and the carbon cycle's response to temperature changes.

On its hand, the purpose of our proposal is not to compete, either in complexity or predictability, with the ESMs. The physical scenarios discussed within the context of our model are, from a theoretical perspective, complementary to the ESMs, as they allow for a consistent tracking of the effects of human drivers as well as the bifurcations and chaotic dynamics that ensue.

For the logistic map dynamics considered here, chaotic does not imply unpredictable, but rather refers to the classification of a discrete dynamical system output, mapped by quantifiable parameters, $\psi$ and $\eta$. In the Anthropocene, chaotic patterns arise exclusively from temperature fluctuations, $\psi$, driven by predictable inputs, namely, the PB.
Chaos, in this context, does not imply uncertainty with respect to the need for mitigation and stewardship measures and policies to stabilise the ES.

Of course, the logistic map modelling of the human impact oversimplifies the complex feedbacks and non-linearities of the encompassed  interactions here discussed. Bifurcation patterns, for instance, are strictly connected to the discreteness of the logistic map, and vanishes away for continuous patterns, where the logistic map is converted into a logistic function.
Increasing the number of input parameters typically transforms the bifurcation lines into {\em spread out} distributions of phase-space points, slightly deviating from the equilibrium regime.
Indeed, this is observed, for instance, in the logistic map supported by two PB parameters, $h_1$ and $h_2$, considered here. In this context, one might expect that multi-parameter models would dramatically increase the unpredictability of the system. This could arise from a simultaneous and dramatic increase in the strength of the PB, all driven by the discreteness of the logistic map dynamics. However, this does not seem to be the real-world scenario, since alternative growth functions (cf., for instance, the linear growth described in Refs.~\cite{Bertolami:2018,Bertolami:2019}) do not yield similar outputs. Such a feature can be regarded as a deficiency of our model: since the emergence of bifurcation and chaotic dynamics, in the mathematical sense we have discussed here, is strictly related to the increasing magnitude of the human-driving parameters, $h_i$, our model cannot be applied retrospectively to identify chaotic patterns, which would emerge only with increasing values of $h_i$, subject to all the constraints mentioned above.

By splitting human activities into their multiple components, we have studied the case where only two components followed logistic maps and interacted with each other.
Even for this simple case, we observed the emergence of chaotic behaviour at the equilibrium points of the ES.
This leads to potentially important consequences if at least some components of human activity actually follow logistic maps, which is a reasonable hypothesis, given the physical limitations of the planet-wide system in which we live.

Alternatively, the assumption of an evolution towards a Hothouse Earth rests on anthropogenic forcing overwhelming natural feedbacks over timescales relevant to human activities/existence, much shorter than those of extreme ES events (e.g., massive volcanic outpourings or asteroid impacts).
In fact, ES dynamics under pre-human natural processes, when the ES experienced much hotter conditions, did not result in a stable Hothouse Earth. 
Thus, given the dominance of natural compensatory mechanisms throughout Earth's history, over large timescales, it is unlikely that a Hothouse Earth state could be stabilised through natural causes alone, without anthropogenic mediation.

With these fairly plausible assumptions, a scenario where it may be impossible to predict the future equilibrium state of the ES, in face of the human activities, becomes a possibility that might be accounted for. The implications for designing managing strategies for the ES such as discussed, for instance in Refs.~\cite{Bertolami:2021a,Bertolami:2021b,Bertolami:2024}, may turn out to be quite dramatic as they might critically impair predictions and decisions making. Indeed, this work sets up a general phenomenological approach that can be used to verify and to further investigate the dynamic behaviour of the ES. Furthermore, it shows that a phenomenological analysis based on data about human activities along the lines described above can indicate the nature of the evolution of the ES regarding its stability and our ability to predict its behaviour. 

\section*{CRediT authorship contribution statement}

{\bf A. E. Bernardini:} Writing -- original draft, review \& editing, Validation, Methodology, Investigation, Formal Analysis.
{\bf O. Bertolami:} Writing -- original draft, review \& editing, Supervision, Validation, Methodology, Investigation.
{\bf F. Francisco:} Writing -- original draft, Methodology, Investigation.

\section*{Funding}

The work of AEB is supported by the Brazilian Agencies FAPESP (Grant No. 2023/00392-8 and Grant No. 2020/01976-5, S\~ao Paulo Research Foundation (FAPESP)) and CNPq (Grant No. 301485/2022-4)

\section*{Declaration of competing interest}

The authors declare that they have no known competing financial interests or personal relationships that could have appeared to influence the work reported in this paper.

\section*{Acknowledgment}

AEB thanks to the Brazilian Agencies FAPESP (Grant No. 2023/00392-8 and Grant No. 2020/01976-5, S\~ao Paulo Research Foundation (FAPESP)) and CNPq (Grant No. 301485/2022-4).

\section*{Data Availability}

Data are in the text itself

\end{document}